\documentclass[a4paper, 12pt]{article}
\usepackage[utf8]{inputenc}
\usepackage[english]{babel}
\usepackage{amsmath, amssymb, amsfonts}
\usepackage{fullpage, indentfirst, cite}
\usepackage{epsfig, graphicx, graphics, color}
\usepackage{hyperref, subcaption, cleveref}

\newcommand{\be}{\begin{equation}}
\newcommand{\ee}{\end{equation}}

\def\lsim{\raise0.3ex\hbox{$\;<$\kern-0.75em\raise-1.1ex\hbox{$\sim\;$}}}
\def\gsim{\raise0.3ex\hbox{$\;>$\kern-0.75em\raise-1.1ex\hbox{$\sim\;$}}}

\title{
{\Large Heavy decaying dark matter and IceCube high energy neutrinos}
\date{}
\author{M.~Kachelrie\ss$^1$, O.~E.~Kalashev$^{2,3}$ and M.~Yu.~Kuznetsov$^{2}$\footnote{mkuzn@inr.ac.ru}
\vspace{.2cm}\\
\footnotesize \it $^1$Institutt for fysikk, NTNU, 7491 Trondheim, Norway\\
\footnotesize \it $^2$Institute for Nuclear Research of the Russian Academy of Sciences,\\ 
\footnotesize \it 60th October Anniversary Prospect 7a, 117312 Moscow, Russia\\
\footnotesize \it $^3$Moscow Institute for Physics and Technology,\\
\footnotesize \it Institutskii pereulok 9, 141700 Dolgoprudny, Moscow Region, Russia}}

\begin{document}

\begin{flushright}
INR-TH-2018-012
\end{flushright}

{\let\newpage\relax\maketitle}

\begin{abstract}	
We examine the hypothesis of decaying heavy dark matter (HDM) in the context
of the IceCube highest energy neutrino events and recent limits on the diffuse
flux of high-energy photons. We consider dark matter (DM) particles $X$ of mass
$10^{6}\leq M_X\leq~10^{16}$~GeV decaying on tree level into $X \rightarrow \nu \bar{\nu}$,
$X \rightarrow e^+e^-$ and $X \rightarrow q \bar{q}$. The full simulation of
hadronic and electroweak decay cascades and the subsequent propagation of the
decay products through the interstellar medium allows us to determine the
permitted values of $M_X$. We show that for leptonic decay channels it is
possible to explain the IceCube highest energy neutrino signal without
overproducing high-energy photons for $M_X~\lesssim~5.5 \cdot 10^{7}$~GeV
and $1.5 \cdot 10^{8}~\lesssim~M_X~\lesssim~1.5 \cdot 10^{9}$~GeV, while hadronic
decays contradict the gamma-ray limits for almost the whole range of $M_X$ values
considered. The leptonic hypothesis can be probed by operating and planned gamma-ray
observatories. For instance, the currently upgrading Carpet experiment
will be capable to test a significant part of the remaining parameter window
within one year of observation. 
\end{abstract}


%%%%%%%%%%%%%%%%%%%%%%%%%%%%%%%%%%%%%%%%%%%%%%%%%%%%%%%%%%%%%%%%%%%%%%%%%%%%%%
\section{Introduction}

The first measurement of the high-energy cosmic neutrino flux by the IceCube
experiment~\cite{Aartsen:2013jdh, Aartsen:2014gkd} has stimulated many
theoretical works, examining in detail the possible sources and  production
mechanisms of these high energy neutrinos --- for a recent review see, e.g.\
Ref.~\cite{review2}. Most often extragalactic sources of astrophysical neutrinos
have been discussed as potential neutrino sources, since in this case the
consistency of the observed arrival directions with isotropy can be naturally
explained. However, none of the proposed source models could be confirmed yet.
The authors of Ref.~\cite{Neronov:2018ibl} suggested recently that an excess
of TeV $\gamma$-rays detected in the Fermi-LAT data at large Galactic latitudes
is the counterpart to the neutrino flux seen by IceCube. This would imply
that a significant fraction of the observed neutrino has a Galactic origin.

In this work, we consider the scenario where the highest energy
IceCube events, namely those constituting the extremely high energy (EHE)
neutrino data set~\cite{Aartsen:2016ngq, Aartsen:2017mau}, are originating from the decays of
heavy dark matter (HDM) particles. Since in such a scenario the flux is
dominated by the Galactic component~\cite{Berezinsky:1997hy}, HDM decays
could explain naturally the TeV $\gamma$-ray excess suggested in
Ref.~\cite{Neronov:2018ibl}. Originally proposed in Ref.~\cite{Feldstein:2013kka},
HDM decays as explanation for the IceCube neutrino flux were studied in detail
in several works assuming different particle physics models~\cite{Esmaili:2013gha,
Bhattacharya:2014vwa, Rott:2014kfa, Esmaili:2014rma, Murase:2015gea, Dev:2016qbd,
Cohen:2016uyg, Sui:2018bbh, Bhattacharya:2017jaw, Borah:2017xgm}.
It was shown, that while explanations of the IceCube data using hadronically decaying
HDM are in tension with the diffuse gamma-ray flux limits~\cite{Kuznetsov:2016fjt, Cohen:2016uyg},
models with leptonic decays are not, remaining viable candidates for the
explanation of the neutrino flux. 
In this work we reconsider the allowed range of HDM masses compatible with
the IceCube EHE events, using recently updated limits on the diffuse
gamma-ray flux from the KASCADE collaboration. We also assess the potential of
future diffuse gamma-ray measurements to constrain this explanation further.
The main difference of the present study from the majority of works explaining the IceCube events
with dark matter decay is the consideration of broad dark matter masses range using
the same technical setup.

For masses $\gsim 50$~TeV, dark matter particles were never in thermal
equilibrium. Possible alternative production
processes via gravitational interactions or other nonthermal processes
in the early Universe have been widely discussed
~\cite{Zeldovich:1971mw, Zeldovich:1977, Kofman:1994rk,Khlebnikov:1996zt,
Khlebnikov:1996wr,Berezinsky:1997hy, Kuzmin:1997jua, Kuzmin:1998kk, Chung:1998rq, Chung:1998zb, Kuzmin:1998uv}
(see also~\cite{Khlopov:1987bh, Fargion:1995xs, Gondolo:1991rn}).
Depending on the specific production mechanism~\cite{Kolb:1998ki,
Kuzmin:1998kk, Kuzmin:1999zk,Chung:1998zb, Chung:2004nh, Gorbunov:2012ij},
the mass of the HDM particle could be constrained from cosmology.
However, in this work we do not fix any particular production model
and consider masses in the range $10^{6} \leq M_X \leq~10^{16}$~GeV.
Since annihilation cross sections are bounded by unitarity as
$\sigma_X^{\rm ann.} \sim 1/M_{X}^2$~\cite{Griest:1989wd}, stable $X$ particles
would lead to an undetectable signal in indirect dark matter searches.
Therefore we consider unstable particles $X$, keeping their lifetime $\tau$ as free parameter.
The two main parameters of a HDM particle, its  mass $M_X$ and lifetime $\tau$,
can be constrained comparing the flux of high-energy particles produced by 
decaying HDM particles with observations. Constraints used are the shape of
the cosmic ray spectrum and the anisotropy of the cosmic ray flux~\cite{Kalashev:2008dh, Kalashev:2017ijd, Marzola:2016hyt}, 
gamma-ray flux limits~\cite{Murase:2012xs, Cohen:2016uyg, Aloisio:2015lva, Esmaili:2015xpa, Kalashev:2016cre, Abeysekara:2017jxs}
and neutrino data~\cite{Abbasi:2011eq, Esmaili:2012us, Cohen:2016uyg, Kuznetsov:2016fjt, Bhattacharya:2017jaw, Aartsen:2018mxl}.

This study complements our previous works~\cite{Kalashev:2016cre, Kuznetsov:2016fjt, Kalashev:2017ijd}
on HDM constraints applying high-energy gamma rays, neutrinos and cosmic-ray anisotropy data.
The paper is organized as follows: In Sec.~\ref{decay} we give a brief overview
of the decay of heavy particles and the simulation of the decay cascades. In Sec.~\ref{propagation},
we describe how we model the  propagation of the decay products through the cosmic medium. In Sec.~\ref{constraints}, we
discuss the procedure of constraining the mass and lifetime of HDM particles with gamma-ray and neutrino
data and derive actual and prospective constraints. Finally, we discuss the results in Sec.~\ref{discussion}.

%%%%%%%%%%%%%%%%%%%%%%%%%%%%%%%%%%%%%%%%%%%%%%%%%%%%%%%%%%%%%%%%%%%%%%%%%%%%%
\section{Heavy dark matter decays}
\label{decay}

We consider HDM particles of masses $10^{6}~\leq~M_X~\leq~10^{16}$\,GeV
decaying at tree level into quarks and leptons.  A characteristic feature of the
decays of particles with masses much larger than the electroweak scale, $M_X\gg m_W$,
is the occurrence of an electroweak cascade in addition to the usual QCD
cascade~\cite{Berezinsky:1997sb,Berezinsky:1999yk}. The hadronic decay channels
of such heavy particles including the underlying (SUSY) QCD cascade were studied
in detail in many works, using both Monte Carlo methods~\cite{Berezinsky:2000up,Aloisio:2003xj}
and the numerical evolution of the DGLAP equations~\cite{Berezinsky:1998ed,
Sarkar:2001se,Aloisio:2003xj,Barbot:2002ep, Barbot:2002gt}. These results were then used in more
recent studies like those of Refs.~\cite{Kalashev:2016cre, Kuznetsov:2016fjt,Kalashev:2017ijd},
where both new constraints on $\gamma$-ray fluxes and the neutrino flux
measurements of IceCube were applied. In contrast, the leptonic decay channel
has received much less attention.

For particles with masses up to 10--100\,TeV a large variety of decay and annihilation
channels into standard model particles have been studied in great detail, see for
instance Ref.~\cite{Cirelli:2010xx} and the references therein. In this work it was
shown that among all possible decay channels the hadronic and leptonic ones yield
the softest and the hardest energy spectra, respectively, for both gamma rays and
neutrinos in the final state. Therefore the constraints on HDM parameters that
would be obtained for other decay channels (e.g.\ those related to gauge bosons)
or for their combinations should lie somewhere between the constraints derived
for the leptonic and hadronic decays. In the following, we consider, therefore, only these two options.

%%%%%%%%%%%%%%%%%%%%%%%%%%%%%%%%%%%%%%%%%%%%%%%%%%%%%%%%%%%%%%%%%%%%%%%%%%%%%%%%
\subsection{Hadronic decay channels}
\label{hadronic}

First, we will discuss the hadronic decay channel, $X \rightarrow \bar q q$, where $q$
denotes a quark with arbitrary flavor. Since $M_X\gg m_q$ for all flavors, the energy
spectra of final-state particles are practically independent on the flavor of the
initial quark. For the evolution of the DGLAP equations, we use the numerical code
from Ref.~\cite{Aloisio:2003xj}, where also a detailed description of its theoretical
basis was given. Here we note only a few key points: For $X \rightarrow \bar q q$,
the three main physical phenomena are the perturbative evolution of the QCD cascade
from scales $t=M_X^2$ to $t\sim 1$\,GeV$^2$, the following hadronization of partons
and subsequent decay of unstable hadrons. The impact of electroweak corrections on
the cascade development is negligible compared to other theoretical uncertainties.

To leading order in $\alpha_s$, the total decay spectrum $F^h(x,s)$ at the scale
$t$ is given by the sum of the parton fragmentation functions $D^h_i(x,t)$, where
$x \equiv 2E/M_X$ is the dimensionless energy fraction transferred to the hadron
and $i$ denotes the parton type: $i= \{q,g\}$. The fragmentation functions at
some high scale $t \sim M_X^2$ can be evolved from the experimentally measured
fragmentation functions at low $t$ with the help of the DGLAP equations~\cite{Gribov:1972rt,
Altarelli:1977zs,Dokshitzer:1977sg} (for details see Refs.~\cite{Aloisio:2003xj, Kalashev:2016cre}).

The initial fragmentation functions are taken from  Ref.~\cite{Hirai:2007cx},
parametrized at the scale $M_Z$, averaged over flavors and extrapolated to the
region $10^{-5} \le x \le 1$. We take into account only the contribution of pion
decays and neglect the contribution of other mesons which was estimated in
Ref.~\cite{Aloisio:2003xj} to be of order $10\%$. Finally the spectra of photons,
electrons and neutrinos are determined from the following expressions, respectively,
\be
\frac{dN_{\gamma}}{dx} = 2 \int\limits_x^1 \frac{dz}{z} \: D^{\pi^0}(z) \,,
\ee
\be
\frac{dN_{\nu}}{dx} = 2\: R \int\limits_{x R}^1 \frac{dy}{y} \: D^{\pi^\pm}(y) +
2 \int\limits_x^1 \frac{dz}{z} f_{\nu_i}\left(\frac yz \right)  \: D^{\pi^\pm}(z) \,;
\ee
\be
\frac{dN_{e}}{dx} = 2\: R \int\limits_{x}^1 \frac{dy}{y} \left( \frac 5 3 - 3y^2 + \frac 4 3 y^3 \right)
\int\limits_{\frac x y}^{\frac{x}{ry}} \frac{dz}{z} \: D^{\pi^\pm}(z)
\ee
where $D^{\pi}(x,s) \equiv [D^{\pi}_q(x,s) + D^{\pi}_g(x,s)]$,
$r = (m_\mu/m_\pi)^2 \simeq 0.573$, $R=\frac{1}{1-r}$ and
the functions $f_{\nu_i}(x)$ are taken from Ref.~\cite{Kelner:2006tc}:
$$
f_{\nu_i}(x)=g_{\nu_i}(x)\,\Theta(x-r)+(h^{(1)}_{\nu_i}(x)+h^{(2)}_{\nu_i}(x))\,\Theta(r-x)\,,
$$
$$
g_{\nu_\mu}(x)=\frac{3-2r}{9(1-r)^2}\,\left(9x^2-6\ln x-4x^3-5\right),
$$
$$
h^{(1)}_{\nu_\mu}(x)=\frac{3-2r}{9(1-r)^2}\,\left(9r^2-6\ln r-4r^3-5\right),
$$
$$
h^{(2)}_{\nu_\mu}(x)=\frac{(1+2r)(r-x)}{9r^2}\left[9(r+x)-4(r^2+rx+x^2)\right] \ ,
$$
$$
g_{\nu_e}(x)=\frac{2}{3(1-r)^2}\,\left[(1-x)\,\left(6(1-x)^2 + r(5+5x-4x^2)\right) + 6r\ln x \right],
$$
$$
h^{(1)}_{\nu_e}(x)=\frac{2}{3(1-r)^2}\,\left[(1-r)\,\left(6-7r+11r^2-4r^3\right) + 6r\ln r \right],
$$
$$
h^{(2)}_{\nu_e}(x)=\frac{2(r-x)}{3r^2}\left(7r^2-4r^3+7xr-4xr^2-2x^2-4x^2r\right) \, .
$$

\begin{figure*}[bt]
\begin{minipage}{0.49\linewidth}
\begin{center}
\includegraphics[width=\linewidth]{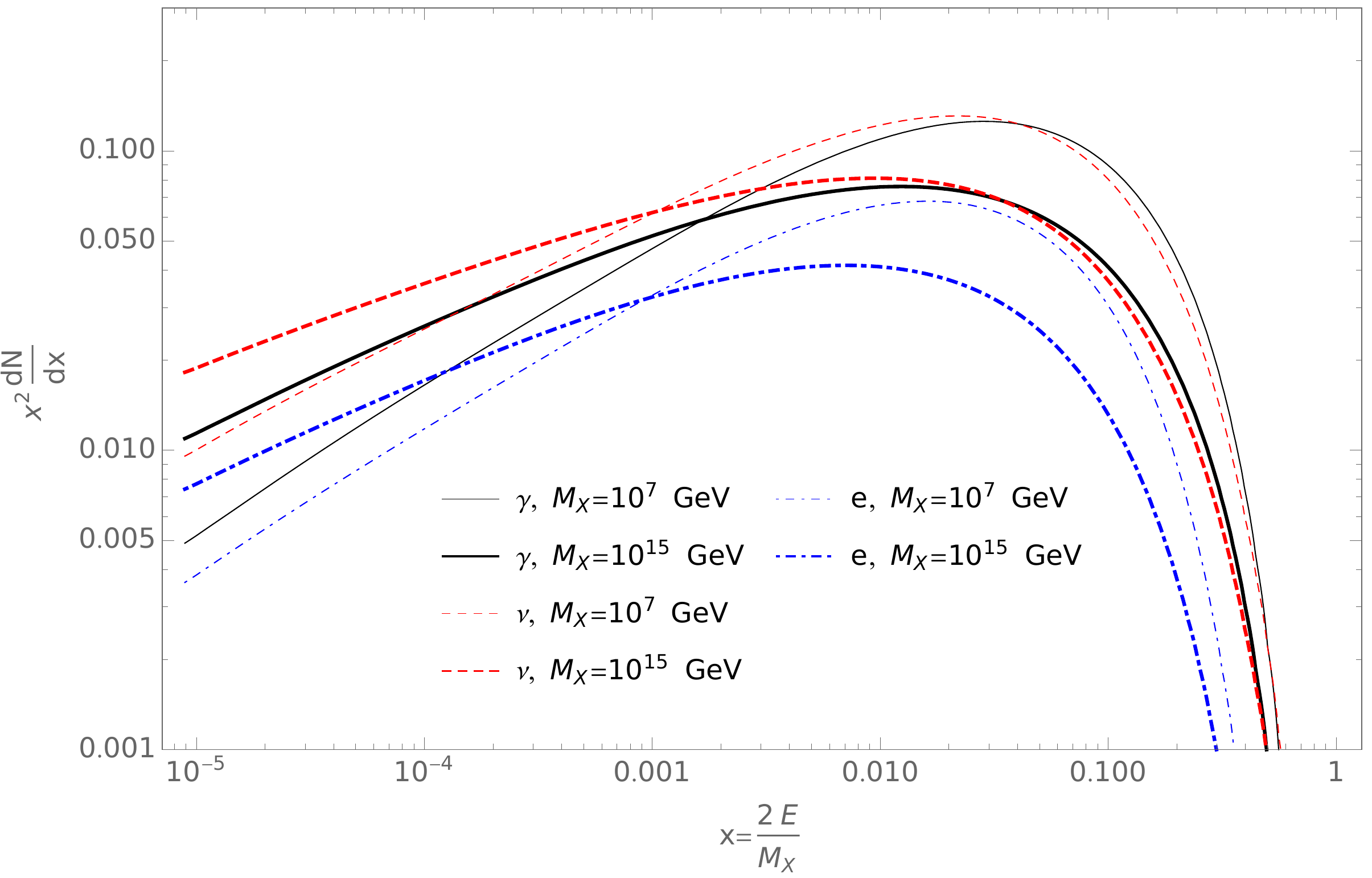}
\subcaption{$X \rightarrow q \bar{q}$}
\end{center}
\end{minipage}
\begin{minipage}{0.49\linewidth}
\begin{center}
\includegraphics[width=\linewidth]{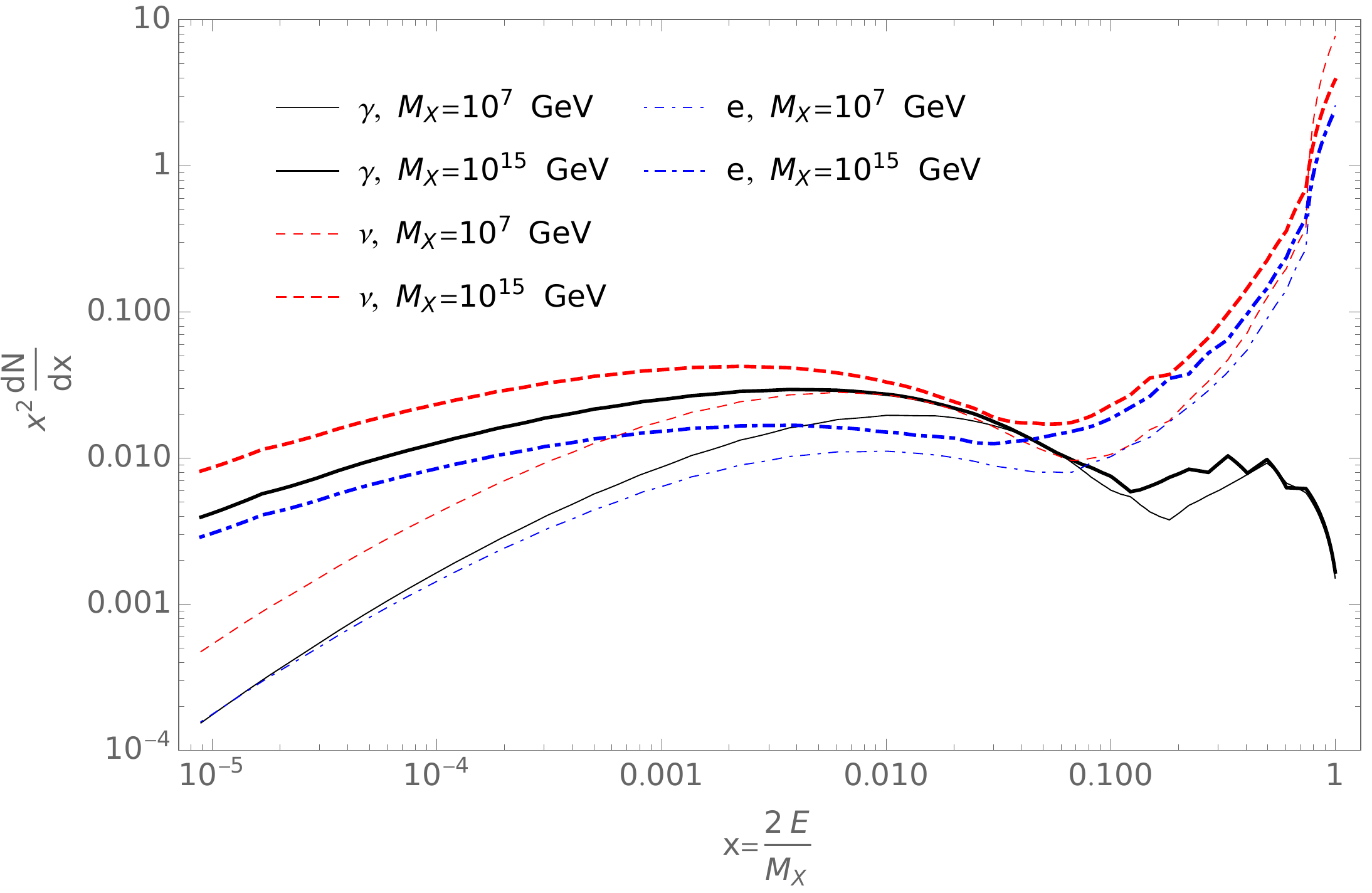}
\subcaption{$X \rightarrow \nu \bar{\nu}$}
\end{center}
\end{minipage}
\caption{
   Prompt spectra of $X$-particle decay for the hadronic channel (left panel) and the
   leptonic channel (right panel) into photons (solid lines), neutrinos (dashed lines)
   and electrons (dot-dashed lines). For each channel, the spectra are shown for two values
   of the mass: $M_X=10^7$~GeV (thin lines) and $M_X=10^{15}$~GeV (thick lines).
}
\label{prompt_spectra}
\end{figure*}

Note that $dN_{e}/dx$ denotes the combined spectrum of electrons and positrons and
$dN_{\nu}/dx$ the combined spectrum of neutrinos and antineutrinos of all flavors.
Also note that the primary electrons contribute to the observable photon flux because of
their interactions in the Galactic halo (see the Sec.~\ref{propagation}).
Examples of photon, electron and neutrino spectra for two values of the $X$ particle
mass and for both hadronic and leptonic decay channels are shown in Fig.~\ref{prompt_spectra}.

%%%%%%%%%%%%%%%%%%%%%%%%%%%%%%%%%%%%%%%%%%%%%%%%%%%%%%%%%%%%%%%%%%%%%%%%%%
\subsection{Leptonic decay channels}

In the case of a tree-level decay to leptons of a particle with mass $M_X\gg m_W$,
large logarithms $\ln^2(M_X^2/m_W^2)$ invalidate perturbation theory, leading to
the development of an electroweak cascade~\cite{Berezinsky:2002hq}. Since the
electroweak gauge bosons split also into quarks $q$, there will be a mutual
transmutation of ``leptonic'' and ``QCD'' cascades. The shape of the hadron
energy spectra is however only marginally influenced by the leptons: first,
because the QCD cascade is determined mainly by gluons $g$, and, second, because
the probability of $q\to q+g$ is much larger than of, e.g., $q\to q+W$.
On the other hand, splittings like $W\to qq$ act continuously as a sink
for the particles and energy of the electroweak part of the cascade. In
order to take into account this effect properly we have performed therefore
a Monte Carlo simulation including both the QCD part as described
in~\cite{Berezinsky:2000up} and the electroweak sector. The latter follows
the scheme described in~\cite{Berezinsky:2002hq}, distinguishing however
now between charged and uncharged leptons and including photons.

The hadronization is based on the procedure described in~\cite{Aloisio:2003xj},
and the resulting photon, electron and neutrino spectra are calculated as in
Sec.~\ref{hadronic}. Similar as the hadronic cascade is insensitive to the flavor
of the initial quark, the leptonic cascade very weakly depends on the choice
of the initial lepton type. The only exception is the spectrum at $x$=1,
corresponding to not branching particles. However, for the mass
range $M_X\geq~10^{6}$\,GeV, these particles give no contribution to the constraints.
To be more specific, the difference between $X\to\bar\nu\nu$ and $X\to e^+e^-$
injection spectra at $x<1$ is within 15\%,; moreover, it decreases with smaller $x$,
while the experimental flux detectability grows with decreasing $x$ making the
highest $x$ parts of the spectrum less relevant for the experimental search.
Therefore it is sufficient to consider the decay $X\to\bar\nu\nu$ as a generic case for the leptonic channel.

%%%%%%%%%%%%%%%%%%%%%%%%%%%%%%%%%%%%%%%%%%%%%%%%%%%%%%%%%%%%%%%%%%%%%%%%%%%%%%%%%%%%
\section{Source distribution and high-energy particles propagation}
\label{propagation}
It is convenient to consider separately the flux from dark matter (DM) decays in the
Milky Way and the extragalactic flux from the entire Universe. 
In this work we use flux predictions for photons with energies above 100 TeV to
build the constraints. The attenuation length of photons with such  energies is
small enough to neglect their extragalactic contribution (see e.g.\ Fig.~7 of Ref.~\cite{Berezinsky:2016feh}).
For the neutrino flux, the contribution of individual distant sources is negligible,
but the total extragalactic flux is sizable because neutrinos propagate over cosmological distances unattenuated.
The Galactic part of the DM decay neutrino flux is described by the following expression:
\be
\label{G_diff_flux}
\frac{d N^{\rm G}}{d E}\left(E\right) = \frac{1}{4\pi \tau M_X} \;
\int\limits_{V} \frac{\rho_{{\rm DM}}\left(R[r]\right)}{4\pi r^2} \frac{d N}{d E}(E, l, b) \; d V \;,
\ee
where $\rho_{DM}\left(R\right)$ is the DM density as a function of the distance $R$ from the Galactic center,
$r$ is the distance from the Earth, $l$ and $b$ are galactic coordinates and
$\frac{dN}{dE}(E, l, b)$ is the spectrum of neutrinos
per decaying $X$ particle. The integration is taken over all the  volume of the
Milky Way halo, for which we assume $R_{\max} = 260$ kpc. We use the Navarro-Frenk-White (NFW) profile
for the dark matter density~\cite{Navarro:1995iw, Navarro:1996gj}~\footnote{In our previous study~\cite{Kalashev:2016cre},
we have compared the DM constraints for the NFW and the Burkert~\cite{Burkert:1995yz}
DM profiles and found that the resulting differences are negligible. Therefore,
we consider here only the NFW profile.} with the parametrization for the Milky Way 
from  Ref.~\cite{Cirelli:2010xx}.

The evaluation of the isotropic extragalactic neutrino flux takes into account the cosmological redshift:
\be
\frac{dN^{\rm EG}}{dE}\left(E_\nu\right) = \frac{1}{4\pi M_X \tau} \int\limits_0^\infty
\frac{ \rho_0 \, c/H_0 }{\sqrt{\Omega_m (1+z)^3 + (1-\Omega_m)}}\frac{dN}{dE}\left[E (1+z) \right] dz
\ee
where $c/H_0 = 1.37 \cdot 10^{28}$ cm is the Hubble radius, $\rho_0 = 1.15 \cdot 10^{-6} \; {\rm GeV}/{\rm cm}^3$ is the present average
cosmological dark matter density, $\Omega_m = 0.308$ and the injected spectrum $\frac{dN}{dE}$
is evaluated as a function of the particle energy at redshift $z$, $E(z) = (1+z)E$.
For neutrinos, the injected flavor composition is also modified by  oscillations during their
propagation. We assume that the flux reaching the Earth is completely mixed, 
i.e.\ that the flavor ratio $\nu_e : \nu_\mu : \nu_\tau$ equals $1 : 1 : 1$.
%It was shown in Ref.~\cite{Bustamante}, that it is a good approximation for
%any combination of neutrino flavors injected.

\begin{figure*}[bt]
\begin{minipage}{0.49\linewidth}
\begin{center}
\includegraphics[width=\linewidth]{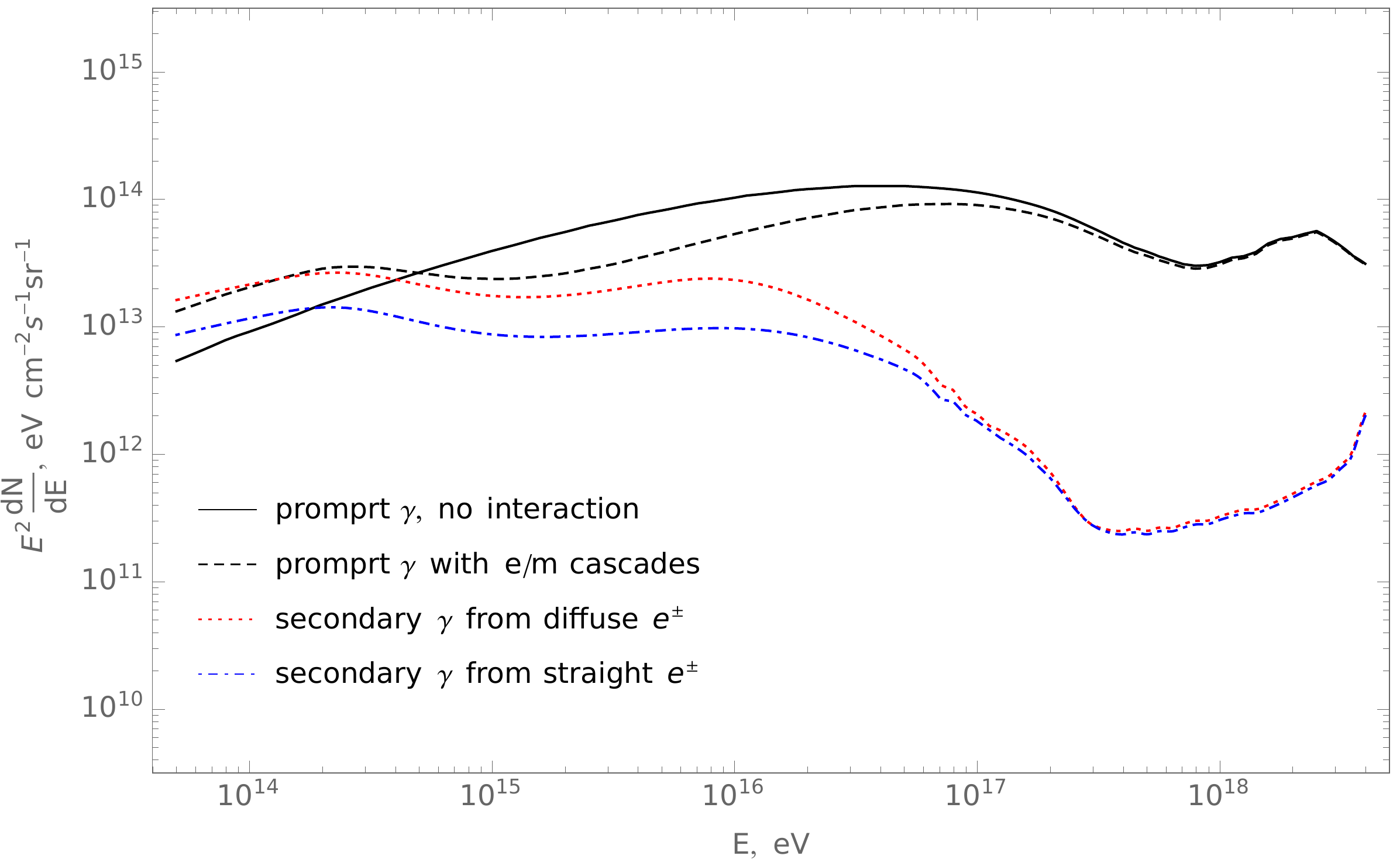}
\subcaption{$X \rightarrow \nu \bar{\nu}$}
\end{center}
\end{minipage}
\begin{minipage}{.05\linewidth}
\end{minipage}
\begin{minipage}{0.49\linewidth}
\begin{center}
\includegraphics[width=\linewidth]{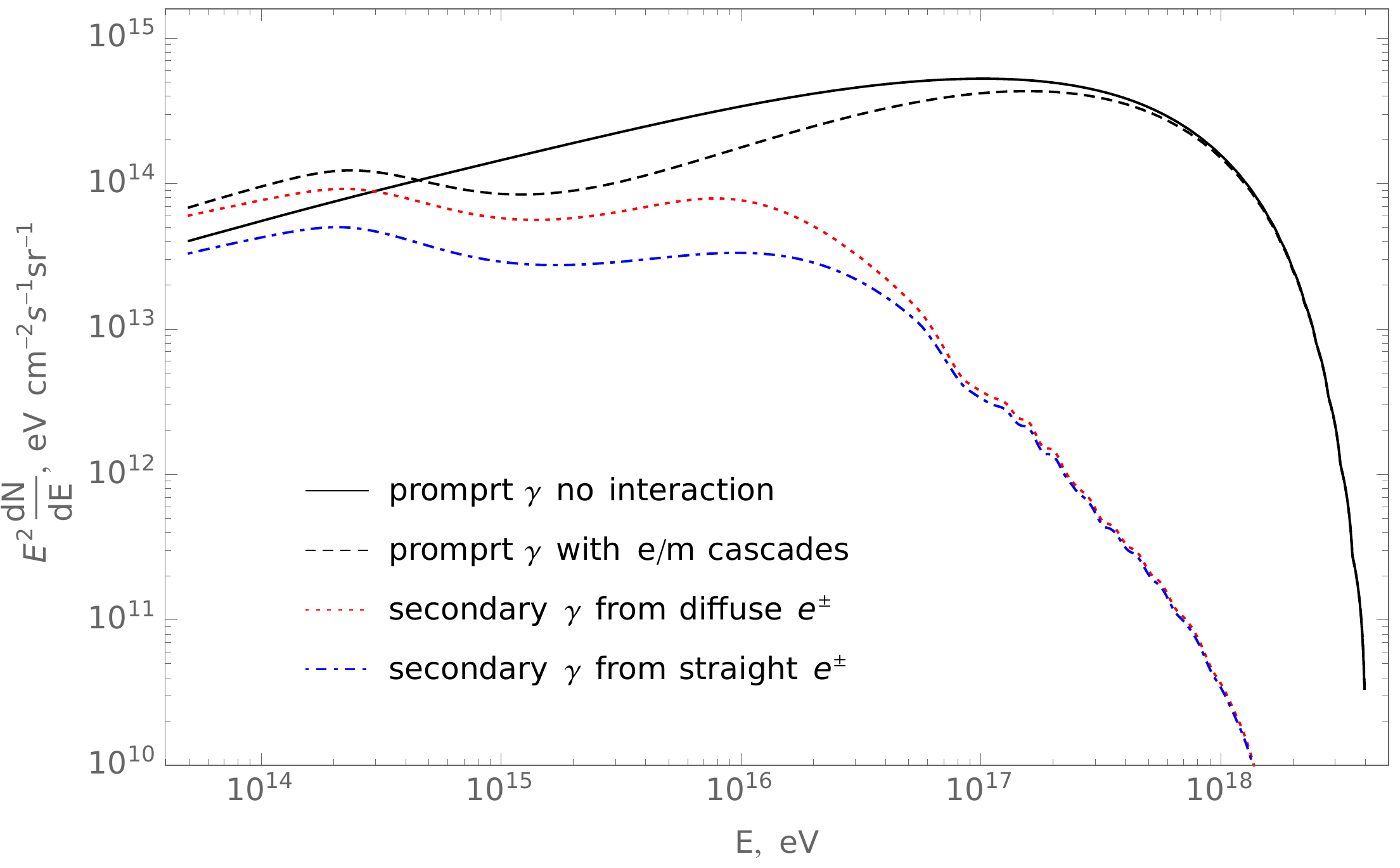}
\subcaption{$X \rightarrow q \bar{q}$}
\end{center}
\end{minipage}
\begin{minipage}{0.49\linewidth}
\begin{center}
\includegraphics[width=\linewidth]{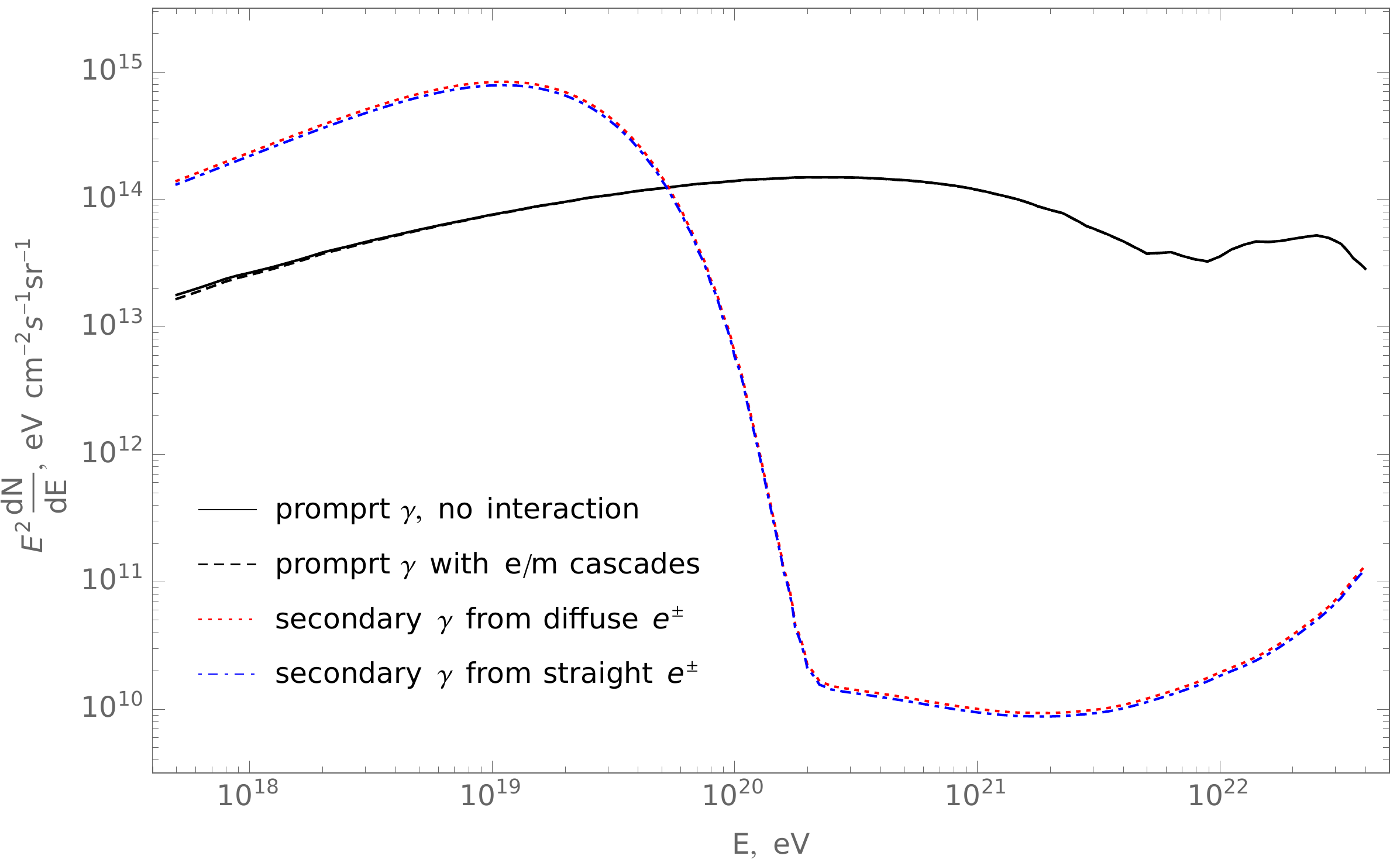}
\subcaption{$X \rightarrow \nu \bar{\nu}$}
\end{center}
\end{minipage}
\begin{minipage}{0.49\linewidth}
\begin{center}
\includegraphics[width=\linewidth]{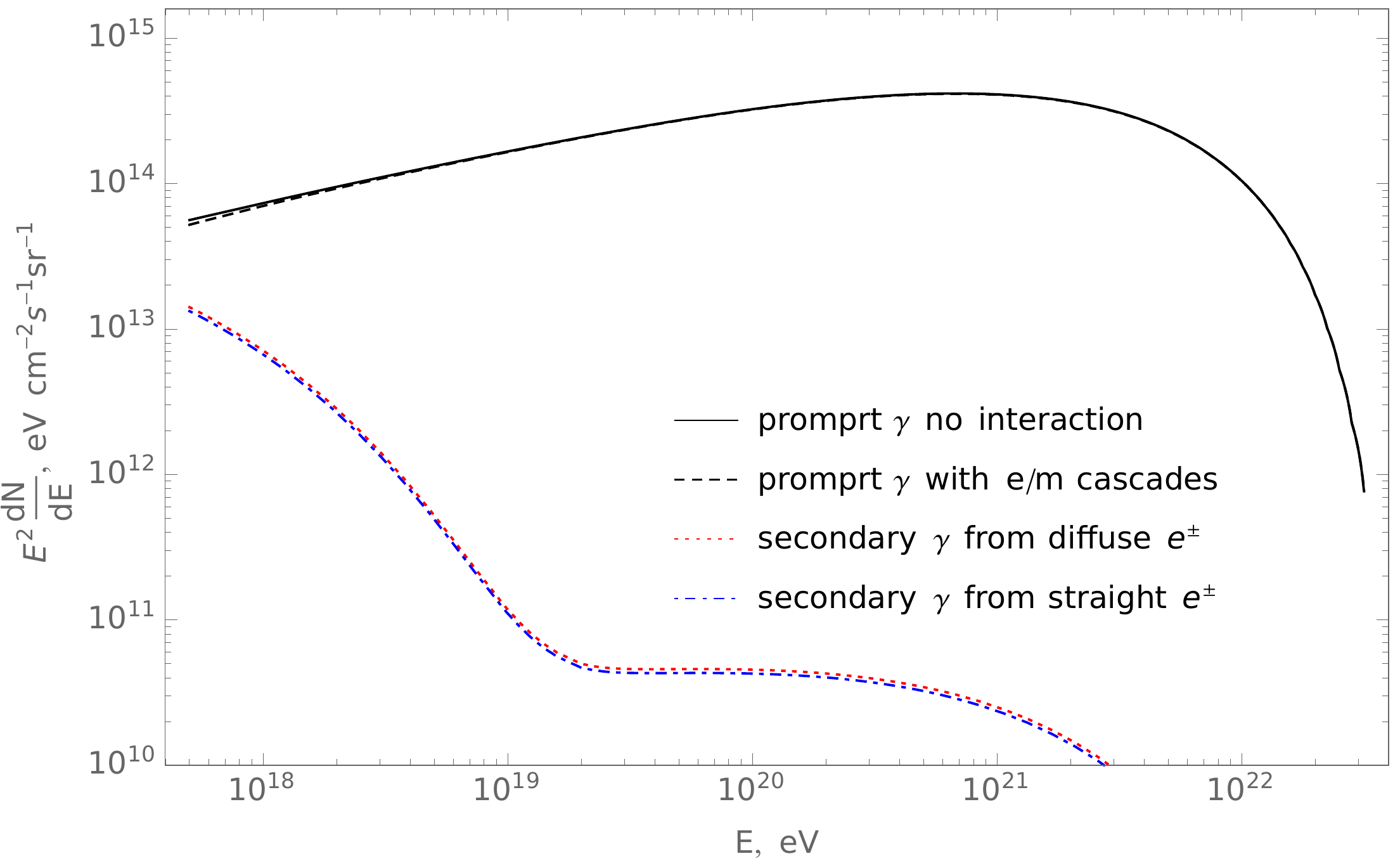}
\subcaption{$X \rightarrow q \bar{q}$}
\end{center}
\end{minipage}
\caption{
Comparison of prompt (noninteracting and with account of the electromagnetic cascades)
and secondary (for the assumptions of straight and diffuse propagation of $e^\pm$) photon
flux from DM decays with mass $M_X=10^{10}$~GeV (top) and $M_X=10^{14}$~GeV (bottom)
in the Milky Way halo (corresponding to the exposure of the Telescope Array experiment).
The magnetic field is assumed to be $B_{\rm halo}=10^{-7}$G.
}
\label{noint_vs_cascade}
\end{figure*}

To calculate the photon flux one has to take into account attenuation effects. $e^\pm$
produced in X-particle decays rapidly loose their energy via synchrotron losses and
up-scattering CMB photons. Both processes contribute to the observable secondary $\gamma$-ray flux.
The $\gamma$-rays with energies above few hundred TeV may in turn produce $e^\pm$ pairs on
CMB photons on a scale of tens of kpc. We take these effects into account by calculating
the flux from  pointlike volumes inside the halo using the numerical
code~\cite{Kalashev:2014xna,Berezinsky:2016feh} and weighting the contributions of different
volumes according to Eq.~(\ref{G_diff_flux}). The numerical code simulates the development
of electron-photon cascades on the CMB driven by the chain of $e^\pm$ pair production and
inverse Compton scattering. While the code allows to calculate the flux of the cascade
and synchrotron photons, it does not take into account deflections of $e^\pm$ by the
magnetic field in the Galactic halo. Since electrons in the code propagate rectilinearly, they produce
less cascade photons. Therefore, the flux of photons calculated in this approximation
should be considered as a conservative lower bound. For the case of $e^\pm$ decay products
an opposite approximation is often used (see e.g.\ Ref.~\cite{Cirelli:2010xx}). Namely,
one  assumes that electrons are kept by magnetic fields in the region of their production
until they loose all their energy via synchrotron or inverse Compton up-scattering of
CMB photons. The secondary gamma rays then propagate towards the observer rectilinearly.
We calculate the secondary $\gamma$-ray flux in both of these approximations, which
we call below straight and diffusive, correspondingly. We can estimate the relevance
of these two approximations as follows:  Electrons with energy $E=10^{15}$~eV have
a Larmor radius $R_L\simeq 100$~pc in a magnetic field of strength 
$B=0.1\mu$G. Assuming that the turbulent field has a coherence length of order $100$\,pc,
electrons with smaller energy diffuse in the large-angle scattering regime.
Then the energy-loss time due to synchrotron radiation is at all energies
of interest smaller than the escape time from an extended magnetic halo of size $\sim 50$~kpc.

The prompt and the secondary photon fluxes obtained in both approximations are compared
with the prompt flux calculated without attenuation effects in Fig.~\ref{noint_vs_cascade}.
One should note that for the hadronic decay channel the secondary $\gamma$-ray flux is
subdominant for both high and low $M_X$: it starts to dominate only at the low $x$ part
of the energy spectra for $M_X \gtrsim 10^{15}$~GeV due to synchrotron emission.
Therefore the impact of the secondary $\gamma$-ray on the HDM constraints is negligible
for the $X \rightarrow q \bar{q}$ channel if $M_X \lesssim 10^{15}$~GeV.

\begin{figure*}[bt]
\begin{minipage}{0.49\linewidth}
\begin{center}
\includegraphics[width=\linewidth]{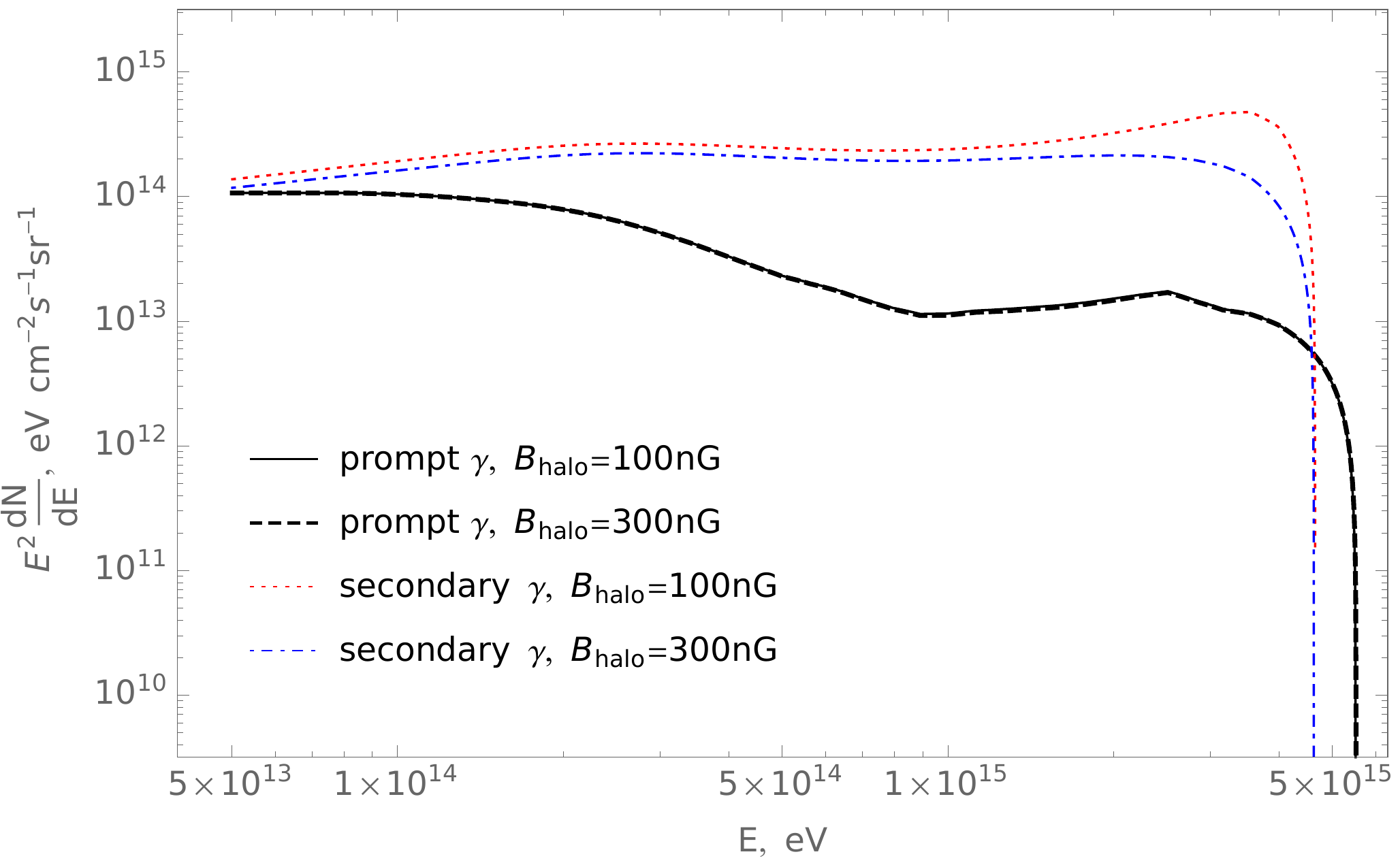}
\subcaption{$M_X = 10^{7}$~GeV, diffuse $e^\pm$ propagation}
\end{center}
\end{minipage}
\begin{minipage}{0.49\linewidth}
\begin{center}
\includegraphics[width=\linewidth]{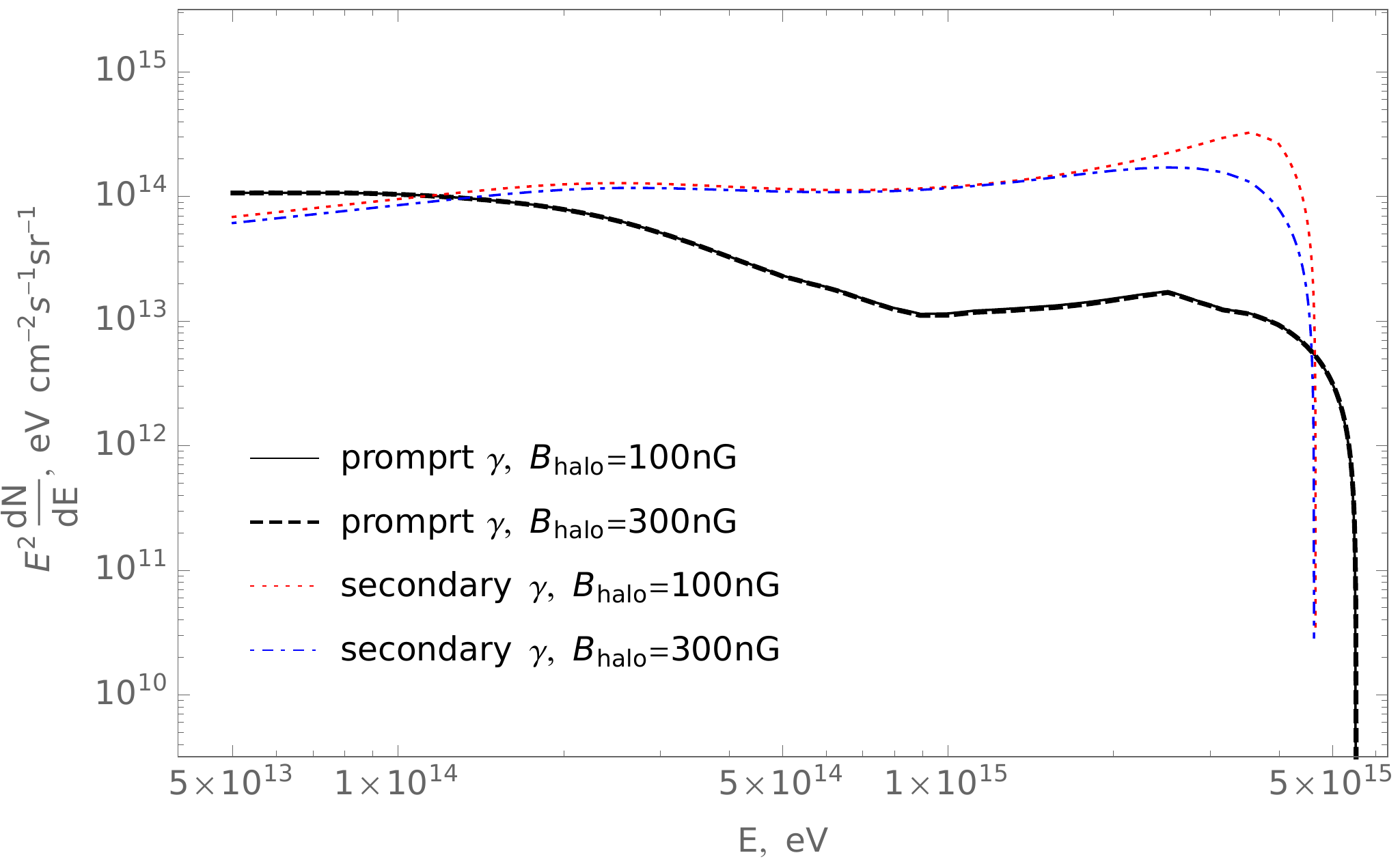}
\subcaption{$M_X = 10^{7}$~GeV, straight $e^\pm$ propagation}
\end{center}
\end{minipage}
\begin{center}
\begin{minipage}{0.49\linewidth}
\includegraphics[width=\linewidth]{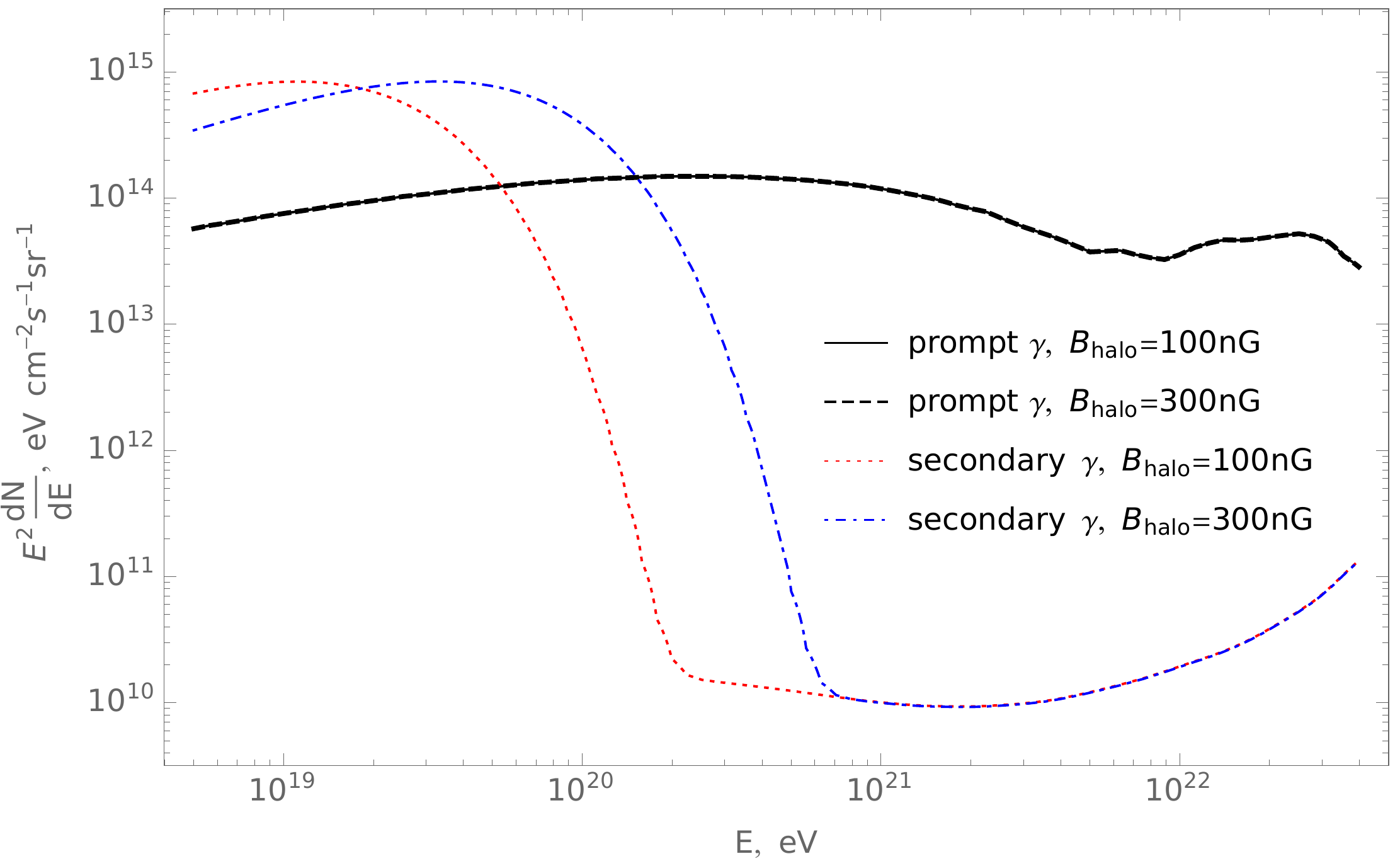}
\subcaption{$M_X = 10^{14}$~GeV}
\end{minipage}
\end{center}
\caption{
Comparison of the prompt and secondary photon fluxes from $X \rightarrow \nu \bar{\nu}$ decays
in the Milky Way halo calculated for two values of galactic halo magnetic field:
$B_{\rm halo} = 10^{-7}$G and $B_{\rm halo} = 3 \cdot 10^{-7}$G. The spectra for
$M_X=10^{7}$~GeV are shown for the assumption of diffuse (top left) and straight (top right)
propagation of prompt $e^\pm$. The spectra for $M_X=10^{14}$~GeV (bottom) are independent of
$e^\pm$ propagation assumption. All spectra are corresponding to the exposure of the
Telescope Array experiment (see Sec.~\ref{constraints} for details).
}
\label{B_comparison_spectra}
\end{figure*}

Besides the $e^\pm$ propagation regime, the secondary gamma-ray flux also depends significantly
on the value of the magnetic field in the Milky Way halo. Both the strength and the structure
of this field are rather uncertain, but, e.g., in the simulations of Ref.~\cite{2012MNRAS.422.2152B}
the field strength is of the order $B\simeq 10^{-7}$G at 50\,kpc distance from the center of
Milky Way-like galaxies. To account for these uncertainties, we consider as two representative
cases the constant values $B=10^{-7}$G and $B=3 \cdot 10^{-7}$G for the magnetic field
strength disregarding its possible dependence on distance from center.
We compare the propagated prompt and secondary spectra for the above cases and
two values of the DM mass in Fig.~\ref{B_comparison_spectra}.

Since the Galactic flux is anisotropic due to the Sun's position in the Galaxy
and propagation effects, the flux prediction for a specific  experiment has to be convolved with its exposure.

\section{Constraints on HDM parameters}
\label{constraints}
In this study we assume that the whole high-energy gamma-ray and neutrino flux is saturated by HDM decays.
The procedure of building model constraints is somewhat different for gamma-ray
and neutrino data. The gamma-ray exposure of EAS observatories
has a universal angular dependence for all gamma-ray energies (since
high energy gamma rays always produce an EAS in the atmosphere). Therefore the
experimental gamma-ray results are given in the form of flux limits averaged over
the experiment field-of-view (FOV). For the neutrino experiments, the exposure as
a function of energy varies for different angular regions; therefore, the data are given as the number of observed events.

\subsection{Gamma-ray constraints}
We first discuss the gamma-ray limits and constraints.
To compare the simulated photon flux with observations we need to convolve it
with the exposure of the given experiment. For EAS experiments with
100\% duty cycle the effective exposure is uniform over right ascension
and sidereal time and can therefore be averaged over these variables. The relative
exposure is given by~\cite{Sommers:2000us}:
\be
\label{exposure}
\omega(a_0,\delta,\theta_\text{max}) \sim (\cos a_0\,\cos\delta\,\sin\alpha_m+\alpha_m\sin a_0\,\sin\delta),
\ee
where $\delta$ is the declination, $a_0$ is the geographical latitude
of the experiment, $\theta_{max}$ is the maximum photon search
zenith angle in the experiment and $\alpha_m$ is given by
\be
\alpha_m=\begin{cases}
0 & ;\xi>1,\\
\pi & ;\xi<-1,\\
\arccos\xi & ; -1 < \xi < 1\,;
\end{cases}
\ee
\be
\xi = \frac{(\cos\theta_\text{max}-\sin a_0\,\sin\delta)}{\cos a_0\,\cos\delta}.
\ee
There is no need to know the overall normalization of the exposure,
since experimental results are given as a flux limit differential over time,
area and FOV. In the class of 100\% duty cycle EAS experiments the most recent results
are given by the Pierre Auger surface detectors~\cite{Aab:2015bza},
the Telescope Array surface detectors~\cite{Rubtsov:2017icrc}, Yakutsk~\cite{Glushkov:2009tn}, 
EAS-MSU~\cite{Fomin:2017ypo}, KASCADE and KASCADE-Grande experiments~\cite{Apel:2017ocm}.
We also use the results of the CASA-MIA experiment~\cite{Chantell:1997gs}.
It is worth noting that some of KASCADE and KASCADE-Grande's recent limits are somewhat less
strict than their previous results. Therefore one expects  weaker
HDM-lifetime constraints. There are also strong
limits provided by the Pierre Auger experiment in the hybrid mode~\cite{Aab:2016agp},
in which duty cycle is however not 100\% and its exposure is nonuniform over
right ascension and sidereal time. Since  there is no publicly available description of
the exposure dependence on sidereal time we use the same formula~(\ref{exposure})
for the Auger hybrid exposure implying the respective HDM constraint is a rough overestimation of the real one.

The integral photon flux received by a given EAS observatory is expressed as:
\begin{equation}
\label{int_flux}
F(E>E_{min}) = \frac{1}{4\pi M_X \tau_X}
\frac{\int\limits_{E_{\min}}^\infty \int\limits_{V} \frac{\rho(R) \omega(l, b, a_0, \theta_{\max})}{r^2} \frac{dN}{dE}(E, l, b) d V d E}{ 2\pi  \int\limits_{-\frac{\pi}{2}}^{\frac{\pi}{2}} \omega(\delta, a_0, \theta_{\max}) \cos(\delta) d \delta}\; ;
\end{equation}
where $\rho(R)$  is the DM density as a function of the distance $R$ from the Galactic center,
$r$ is the distance from the Earth, $l$ and $b$ are Galactic coordinates and
$\frac{dN}{dE}(E, l, b)$ is the spectrum of primary and secondary photons
produced per decaying $X$ particle. The integration in the numerator
is taken over all the volume of the DM halo  ($R_{\max} = 260$ kpc) and in the denominator
over all sky (the cut of the unseen sky regions is included into
the definition of $\omega$).

\begin{figure*}[bt]
\begin{minipage}{0.49\linewidth}
\begin{center}
\includegraphics[width=\linewidth]{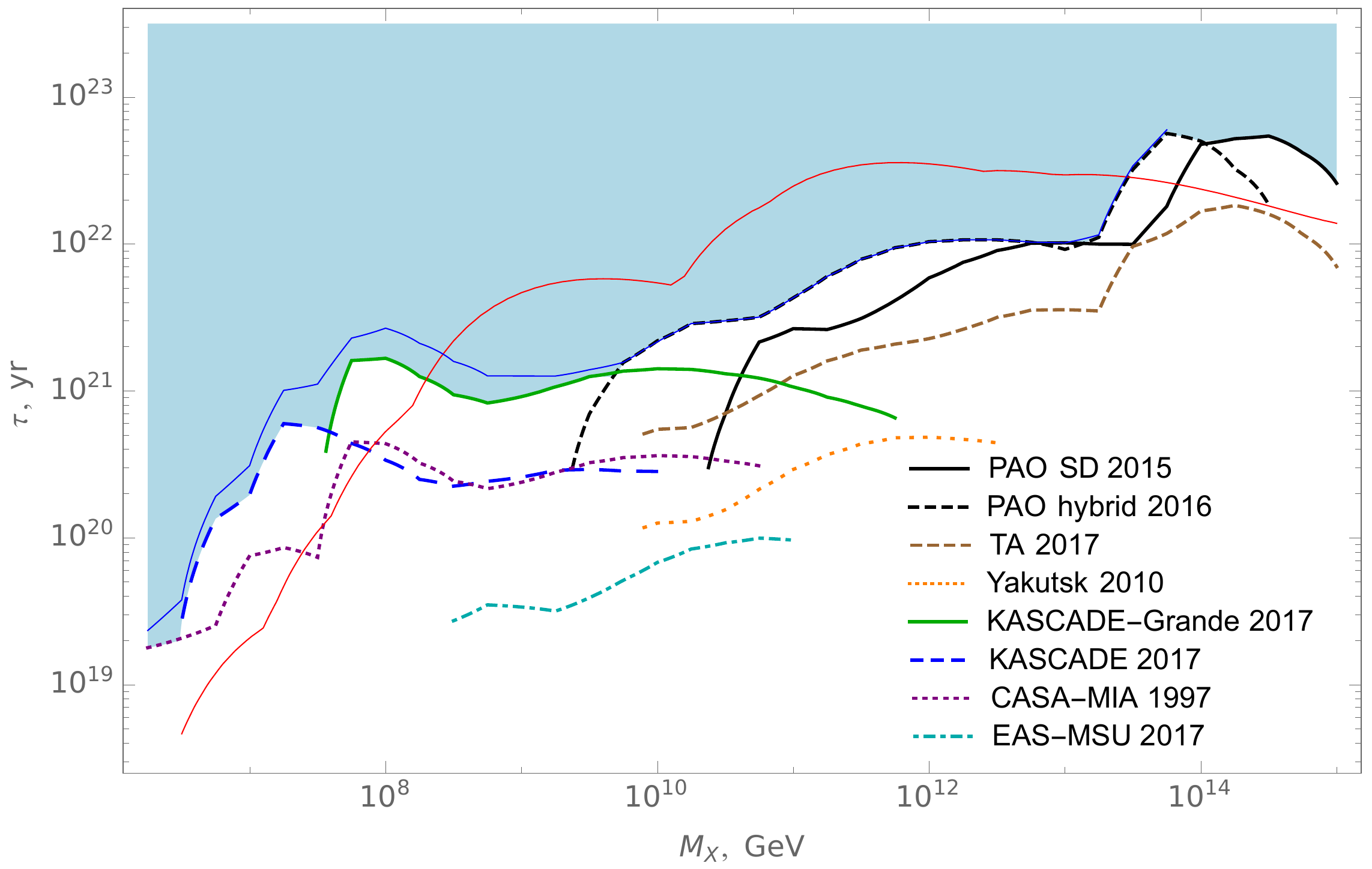}
\subcaption{$\gamma$-constraints}
\end{center}
\end{minipage}
\begin{minipage}{0.49\linewidth}
\begin{center}
\includegraphics[width=\linewidth]{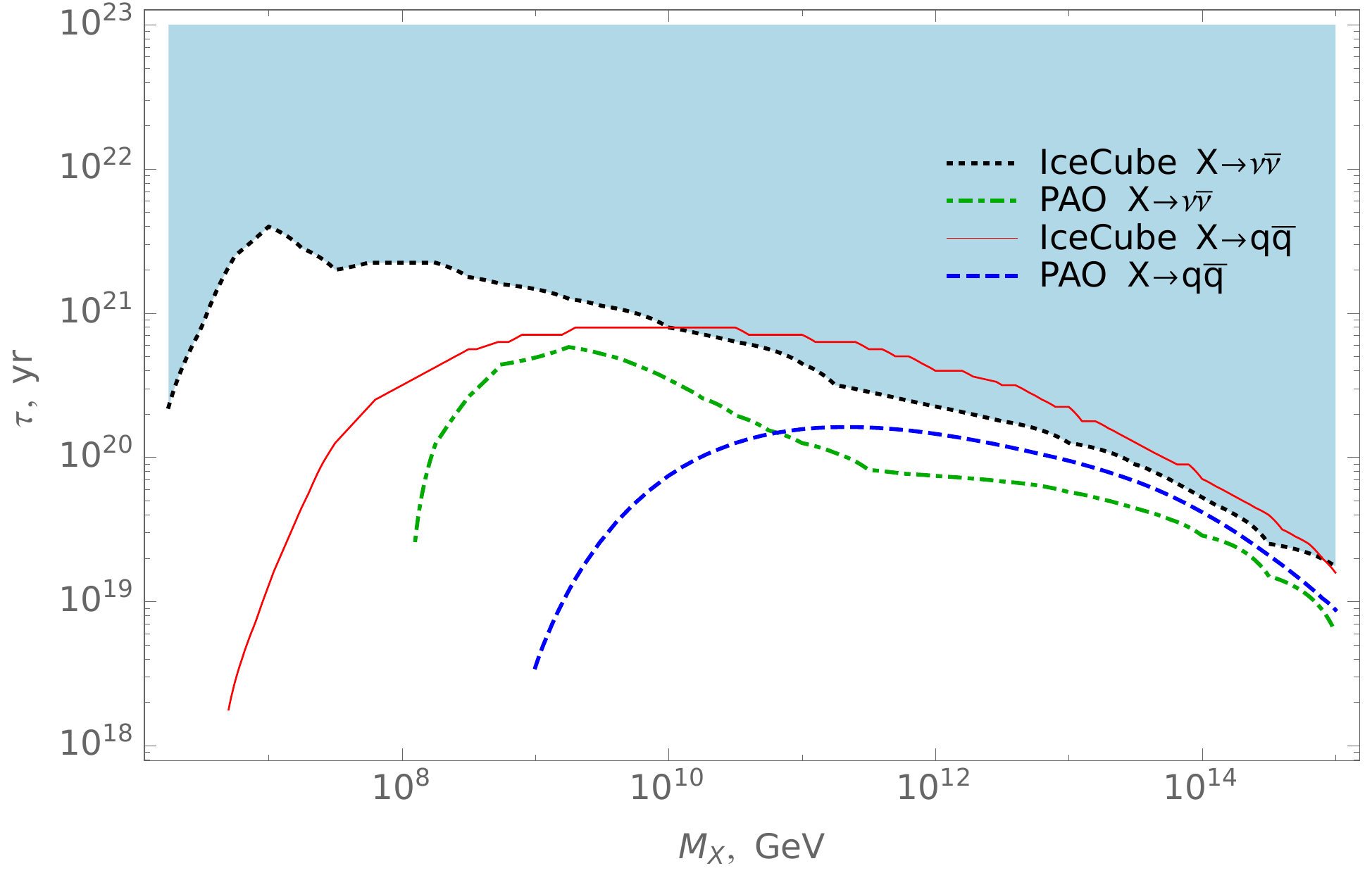}
\subcaption{$\nu$-constraints}
\end{center}
\end{minipage}
\caption{
Constraints on the parameters of HDM for the $X \rightarrow \nu\bar{\nu}$ decay
channel derived from various experimental gamma-ray limits (left panel) and
neutrino data (right panel). The white area is excluded. The gamma constraints
shown assume secondary gamma rays from quasistraight propagation of $e^\pm$
in Galaxy (white area) or diffuse $e^\pm$ propagation (below solid blue line).
The all-experiments constraint from both gamma ray and neutrino for the
$X \rightarrow q\bar{q}$ decay channel is shown by the solid red line.
For neutrino data, the constraints are at 90\% C.L. For gamma-ray data, the KASCADE, CASA-MIA,
KASCADE-Grande and EAS-MSU constraints are at 90\% C.L.; the Yakutsk, Pierre Auger and Telescope Array constraints are at 95\% C.L.
}
\label{nu_exp}
\end{figure*}

Given the predicted gamma-ray flux observable by the particular experiment we obtain the minimal
DM lifetime by varying it until the flux matches at least one of the experimental
limits. Separate constraints for $X \rightarrow q \bar{q}$ and
$X \rightarrow \nu \bar{\nu}$ ($X \rightarrow e^+ e^-$) decay channels
are shown in Fig.~\ref{nu_exp}. For both channels the
most strict constraints are given by the Pierre Auger Observatory, KASCADE-Grande, KASCADE and CASA-MIA.
For each channel we show the constraints for two distinct cases of secondary
photon flux calculation discussed in Sec.~\ref{propagation}: quasistraight
propagation of $e^\pm$ and its diffusion until the complete emission of energy
into photons. One can see that these two cases produce similar constraints
starting from $M_X \gtrsim 5\cdot10^9$~GeV.

\subsection{Neutrino constraints}
In this work we use the neutrino observations  by IceCube~\cite{Aartsen:2016ngq},
recently updated in Ref.~\cite{Aartsen:2017mau} (Sec. 9). This data set contains two
events with 2.6~PeV and 2.7~PeV energies, we assume that they have HDM origin.
For comparison, we also use the results of neutrino nonobservation by Pierre Auger~\cite{Zas:2017xdj}.

We build HDM constraints with the neutrino
data using the procedure discussed in detail in Ref.~\cite{Kuznetsov:2016fjt}.
Namely, we compare the observed number of neutrino events with the number
predicted by the model assuming the exposure of the particular experiment~\cite{Anchordoqui:2002vb}.
For the Galactic neutrino flux the calculation yields:
\be
N_{\rm G}^\nu = \frac{1}{4\pi M_X \tau} \int\limits_{\Delta E} \int\limits_{V} \rho\left[R(r,\delta,\alpha)\right]\:
\omega(E, \delta, \alpha)\: \frac{dN}{dE}(E)\: \cos(\delta)\: dr\: d\delta\: d\alpha\: dE\; ,
\ee
where the integration is performed over all the volume of the dark-matter halo ($R < 260$ kpc)
and over the neutrino energy range $\Delta E$ accessible by a given experiment.
In practice, the exposure is given for several zenith angle bands, averaged over each band.
For IceCube we adopt the exposure as a function of declination
(which uniquely translates to zenith angle in the case of IceCube) and energy as it is
given in Ref.~\cite{Abbasi:2011ji} and normalize it to the actual IceCube exposure
of Ref.~\cite{Aartsen:2016ngq}. For the Pierre Auger Observatory we use the exposure given in Ref.~\cite{Zas:2017xdj}.

The number of events from the extragalactic neutrino flux is
\be
N_{\rm EG} = \int\limits_{\Delta E} \omega(E)\: \frac{dN_{\rm EG}}{dE}(E)\: dE\; ,
\ee
where the exposure $\varepsilon(E)$ is integrated over the celestial sphere.
The total number of events predicted by the model is
\be
\label{N_total}
N_{\rm th} = N_{\rm G} + N_{\rm EG}\, .
\ee

For each mass $M_X$ we constrain the lifetime $\tau$ according to the binned
Poisson Monte Carlo procedure described in Ref.~\cite{Kuznetsov:2016fjt}.
Namely, we generate many Monte--Carlo sets with
the number of events in each energy bin, $N_{\rm MC}^i$, following a Poisson
distribution with mean $\lambda^i = N_{\rm th}^i$, which equals the expected
average number of events for a given $\tau$ and $M_X$. The minimal allowed
value of $\tau$ is the value for which the fraction of Monte Carlo sets
with $N_{\rm MC}^i > N_{\rm obs}^i$ in at least one bin reaches the given confidence level.
The method is applicable when the number of background events is negligible, which is
true for the discussed IceCube data set~\cite{Aartsen:2016ngq}.
The constraints on the parameter space $\{M_X, \tau\}$ are presented in Fig.~\ref{nu_exp}.

\section{Discussion}
\label{discussion}

\begin{figure*}[bt]
\begin{minipage}{0.49\linewidth}
\begin{center}
\includegraphics[width=\linewidth]{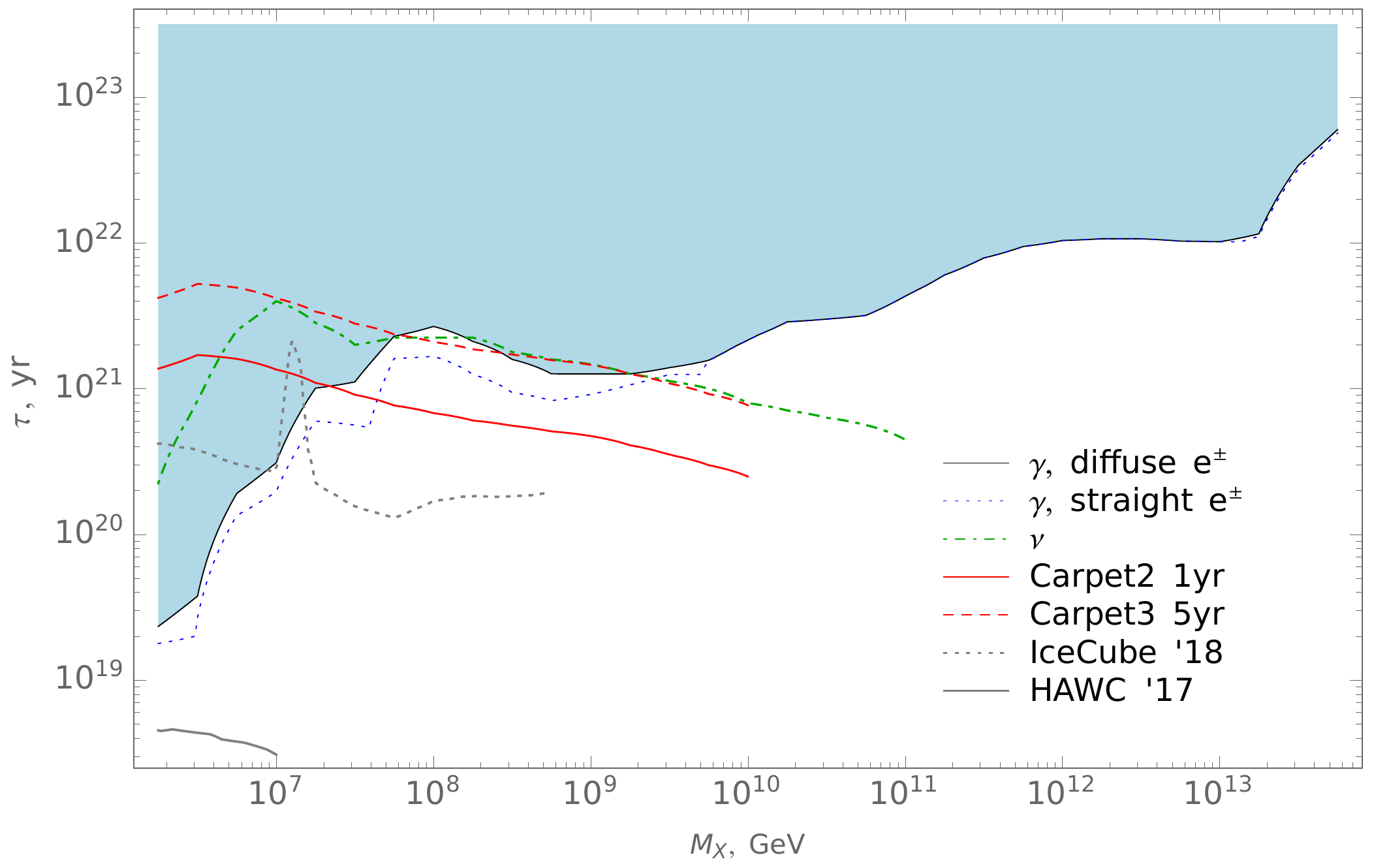}
\subcaption{$X \rightarrow \nu\bar{\nu}$}
\end{center}
\end{minipage}
\begin{minipage}{0.49\linewidth}
\begin{center}
\includegraphics[width=\linewidth]{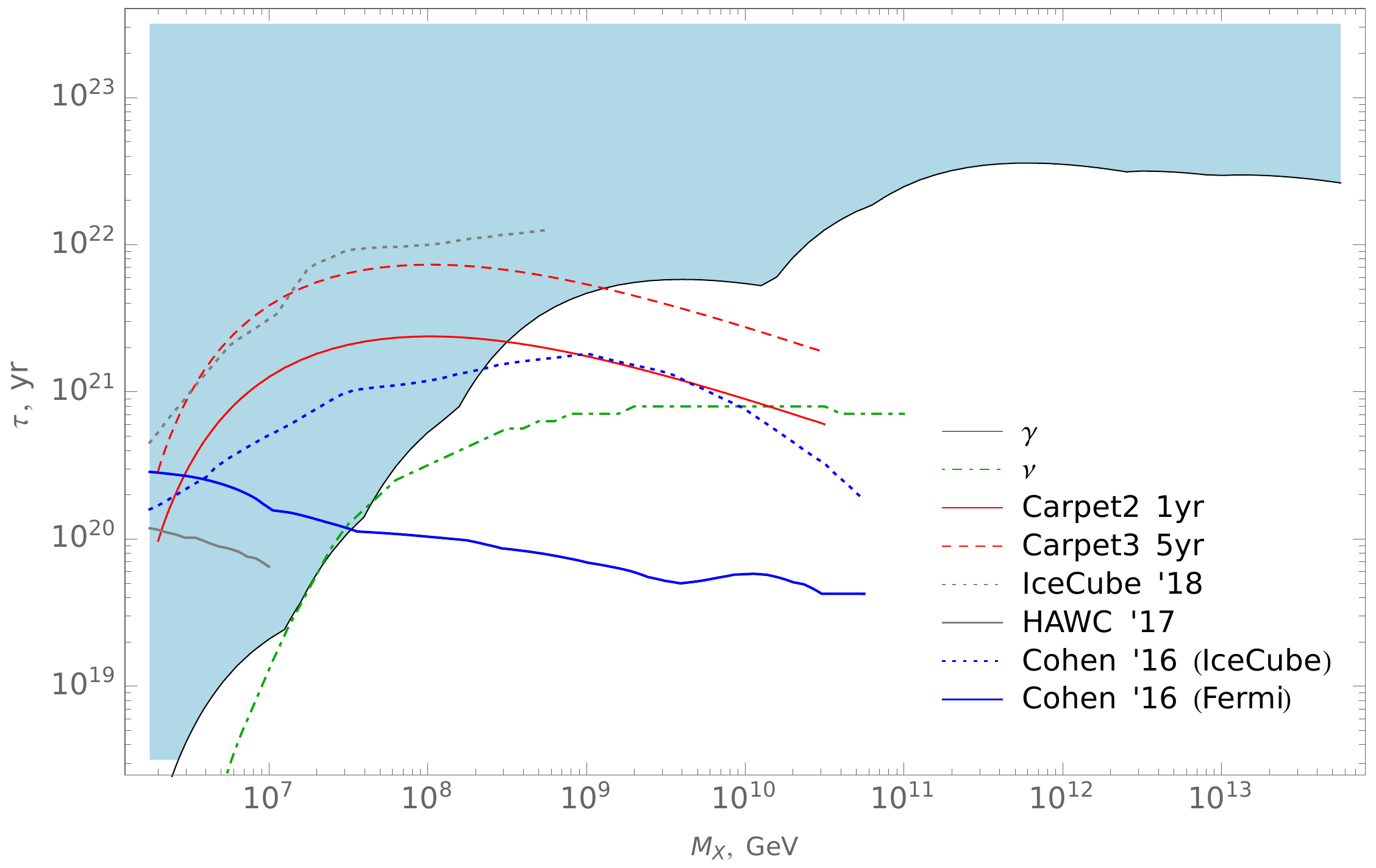}
\subcaption{$X \rightarrow q \bar{q}$}
\end{center}
\end{minipage}
\caption{
Gamma ray vs. neutrino constraints on HDM parameters for leptonic decay (left panel) and
hadronic decay channels (right panel). For the leptonic decay, the constraints with
secondary gamma rays from straight $e^\pm$ (blue dashed line) and diffuse $e^\pm$
(black solid line) are shown. For both decay channels, the prospective constraints
for diffuse gamma-ray searches of the Carpet experiment~\cite{Dzhappuev:2015hxl}
are shown for 1 year of Carpet2 observations (solid red line) and 5 years of Carpet3
observations (dashed red line). The constraints for $X \rightarrow \nu\bar{\nu}$ channel
from Refs.~\cite{Aartsen:2018mxl, Abeysekara:2017jxs} and for the $X \rightarrow b\bar{b}$
channel from Refs.~\cite{Aartsen:2018mxl, Cohen:2016uyg, Abeysekara:2017jxs} are
shown for comparison on the left panel and right panel, respectively.
}
\label{nu_vs_gamma}
\end{figure*}
We have studied the hypothesis of heavy dark matter particles $X$ of mass $10^{6}\leq M_X\leq~10^{16}$~GeV
decaying on tree level into $X \rightarrow \nu \bar{\nu}$, $X \rightarrow e^+e^-$
and $X \rightarrow q \bar{q}$ in the context of the IceCube highest energy neutrino events and recent limits on the diffuse
flux of high-energy photons. For neutrino flux constraints we have selected the data
set~\cite{Aartsen:2016ngq, Aartsen:2017mau} among the various IceCube measurements as the
most conservative one. The analysis cuts for this set are relatively strict, aiming at
eliminating all atmospheric neutrino backgrounds. This should lead, of course, to a
reduction in the total neutrino exposure, but two events with 2.6~PeV and 2.7~PeV
energies in the resulting neutrino set are found to be consistent with astrophysical
neutrino comparing to Monte Carlo simulations. In this study we assume that these
events are of HDM decay origin, which does not contradict the mentioned  IceCube studies.
There are several works~\cite{Cohen:2016uyg, Sui:2018bbh, Bhattacharya:2017jaw}
that use other IceCube data sets~\cite{Aartsen:2015knd, Aartsen:2017mau}
to constrain the HDM parameters, as well as the recent work on HDM search by the IceCube
collaboration itself~\cite{Aartsen:2018mxl}. These studies set  HDM constraints
employing a likelihood analysis of the neutrino spectrum, assuming an arbitrary
combination of neutrinos with astrophysical and HDM decay origin in a wide energy range.
The larger exposure of the data sets and the complexity of the fit models used in these studies results
in  quite strong constraints on HDM parameters. In contrast,
the constraints derived in this work with a relatively simple approach are more conservative.

A comparison of gamma ray and neutrino constraints 
for the $X~\rightarrow~\nu\bar{\nu}$ and $X~\rightarrow~q\bar{q}$ decay channels is shown
in Fig.~\ref{nu_vs_gamma} together with some $\gamma$-ray and neutrino constraints
from other studies. One should note that for hadronically decaying HDM the neutrino
constraints obtained in this work are weaker than the gamma-ray ones for almost the entire
mass range considered. This implies that hadronically decaying HDM as an explanation of
the highest energies IceCube events is disfavored. For the leptonic decay channel the situation is the opposite:
the IceCube high energy signal could be explained by HDM with masses up to $M_X~\lesssim~5.5 \cdot 10^{7}$~GeV
and $~1.5 \cdot 10^{8}~\lesssim M_X~\lesssim~1.5 \cdot 10^{9}$~GeV.

The part of parameter space allowed by the current $\gamma$-ray limits can be further constrained
with ongoing and future high-energy gamma experiments. For instance the Carpet EAS experiment of
the Baksan Neutrino Observatory is currently operating and undergoing consecutive upgrades to an area of
410 m$^2$ (Carpet2) and 615 m$^2$ (Carpet3). After the later upgrade it will be sensitive to $\gamma$-rays
in the energy range from 100~TeV to 30~PeV, with more than an order of magnitude increased sensitivity at 100~TeV
compared to the current KASCADE limit~\cite{Dzhappuev:2015hxl}. In Fig.~\ref{nu_vs_gamma} we illustrate the range of model
parameters which can be excluded by the Carpet experiment, if the $\gamma$-ray flux is not detected.
One can see that the experiment will be capable to cover a significant part of the remaining open
parameter space of models explaining the IceCube neutrino flux within one year of observation.
It is able to close this window practically within five years of observation after the second upgrade.

\begin{figure*}[bt]
\begin{minipage}{0.49\linewidth}
\begin{center}
\includegraphics[width=\linewidth]{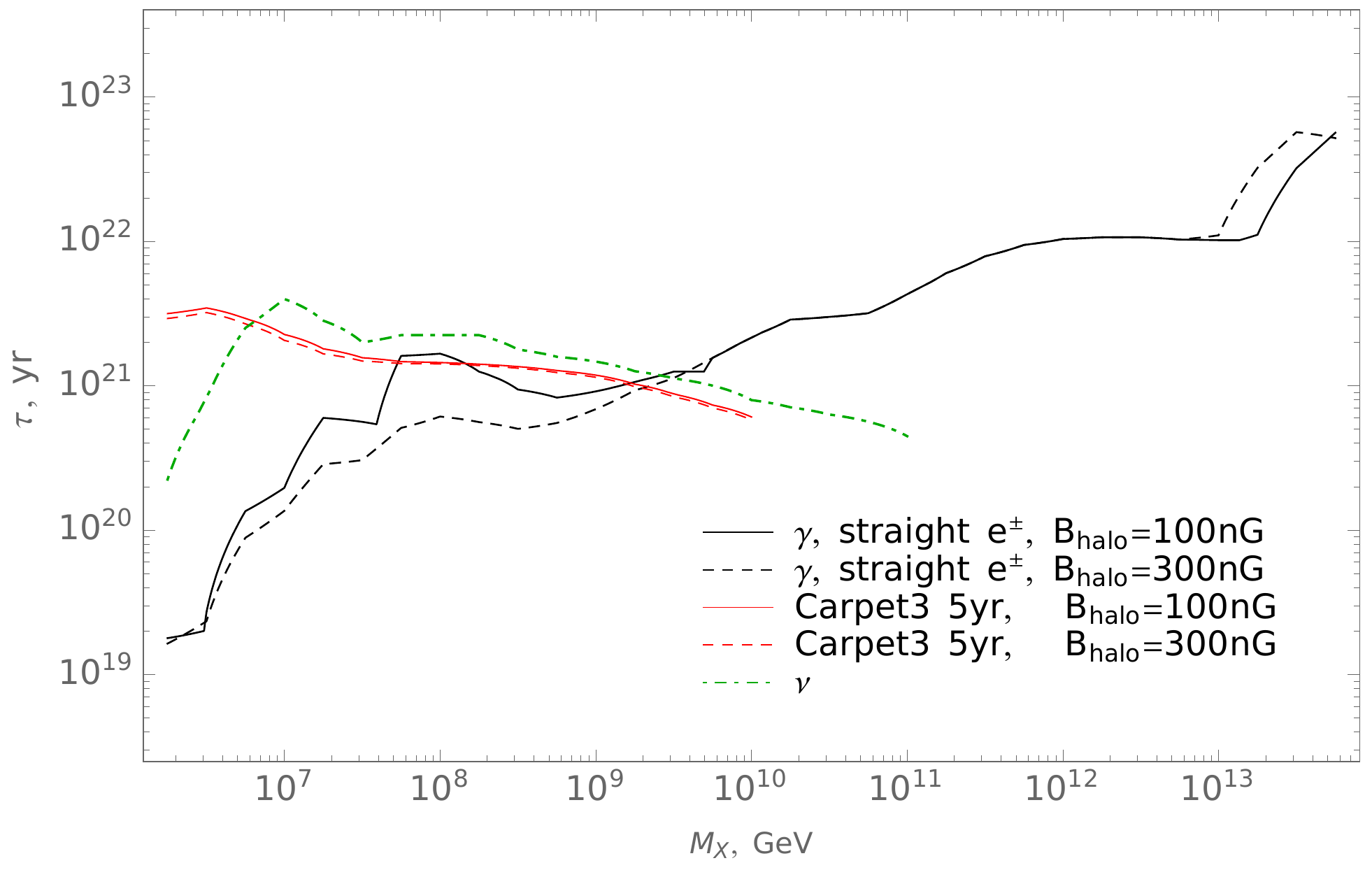}
\subcaption{Straight $e^\pm$ propagation}
\end{center}
\end{minipage}
\begin{minipage}{0.49\linewidth}
\begin{center}
\includegraphics[width=\linewidth]{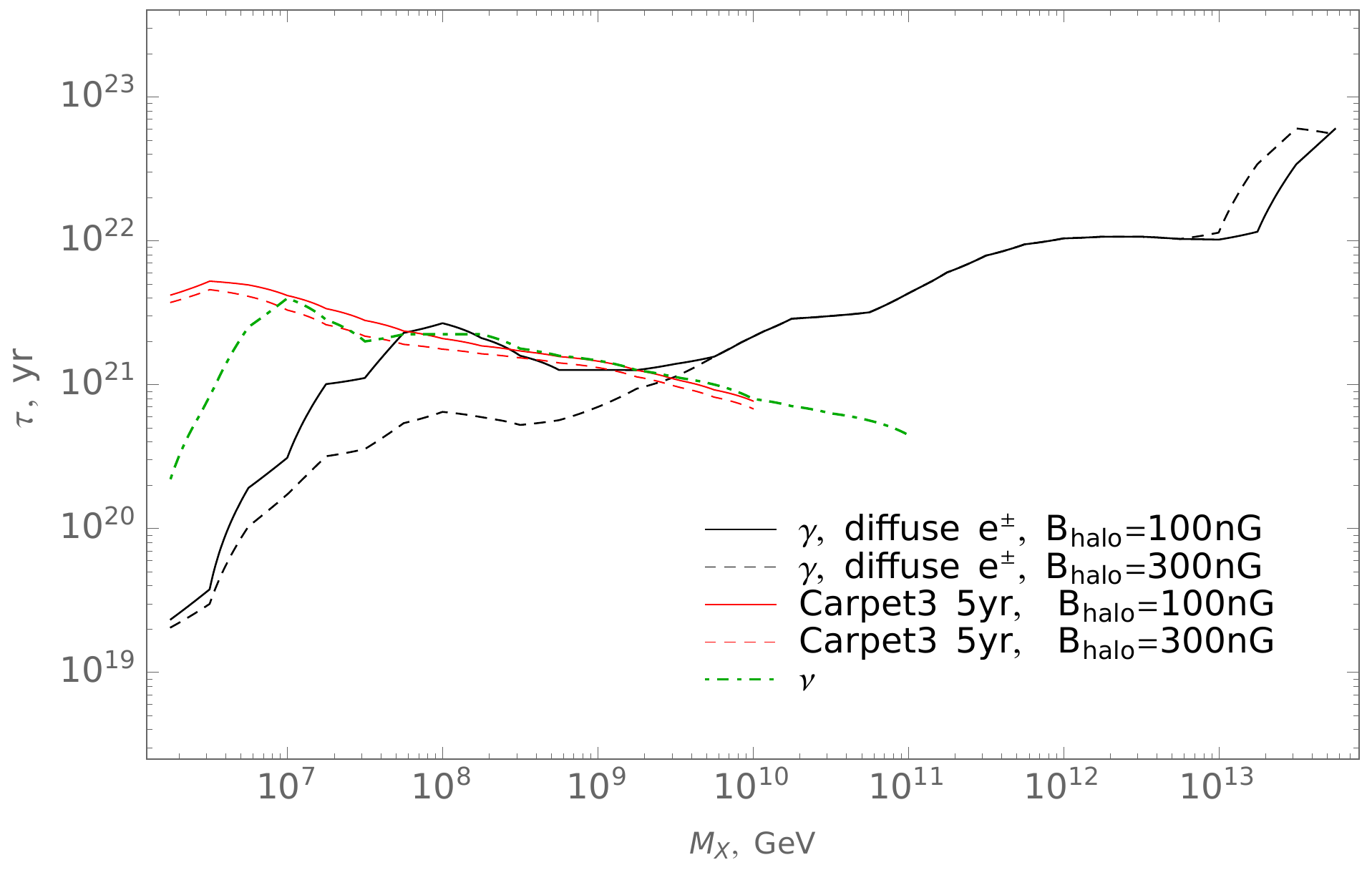}
\subcaption{Diffuse $e^\pm$ propagation}
\end{center}
\end{minipage}
\caption{
Gamma-ray vs. neutrino constraints on HDM parameters for leptonic decay: comparison of
$B_{\rm halo} = 10^{-7}$G and $B_{\rm halo} = 3 \cdot 10^{-7}$G cases for straight (left panel)
and diffuse (right panel) propagation of electrons. In both cases the prospective constraints for
diffuse gamma-ray searches of the Carpet experiment~\cite{Dzhappuev:2015hxl} are also shown by red lines.
}
\label{B_comparison}
\end{figure*}

However, the constraint obtained for the $X~\rightarrow~\nu\bar{\nu}$ channel is dependent
on the assumed value of the halo magnetic field. We compare the limits derived using the
field strength values $B_{\rm halo} = 10^{-7}$G and $B_{h\rm alo} = 3 \cdot 10^{-7}$G
in Fig.~\ref{B_comparison}: For higher magnetic field strength the flux of secondary photons
is suppressed and the constraint weakens. We  should also stress that in the case of $X~\rightarrow~q\bar{q}$
channel the limits are  not affected by the strength of the magnetic field, since the contribution
of the secondary gamma rays to the total gamma-ray flux is negligible for this channel.

\section*{Acknowledgements}
We would like to thank S.~Troitsky and D.~Semikoz for helpful discussions.
The work of O.~Kalashev and M.~Kuznetsov is supported by
the Foundation for the Advancement of Theoretical Physics and Mathematics ``BASIS''.

\suppressfloats

%%%%%%%%%%%   bibliography   %%%%%%%%%%%%%%%%
\bibliographystyle{unsrturl}
\bibliography{ref,ref2,ref3}

\end{document}